\documentclass[a4paper,fleqn]{article}

\usepackage[utf8]{inputenc}
\usepackage[T1]{fontenc}

\DeclareUnicodeCharacter{2212}{-}
\DeclareUnicodeCharacter{2013}{--}
\DeclareUnicodeCharacter{2014}{---}
\DeclareUnicodeCharacter{2019}{'}
\DeclareUnicodeCharacter{202F}{\,}

\usepackage[margin=0.9in]{geometry}

\usepackage[authoryear]{natbib}

\usepackage{amsmath}
\usepackage{amssymb}
\usepackage{amsfonts}
\usepackage{mathtools}
\usepackage{bm}
\usepackage{slashed}

\numberwithin{equation}{section}

\usepackage{graphicx}
\usepackage{subcaption}
\usepackage{booktabs}
\usepackage{array}
\usepackage{multirow}
\usepackage{placeins}
\usepackage{float}

\usepackage{xcolor}

\definecolor{linkblue}{RGB}{0,70,140}

\usepackage{amsthm}

\usepackage[
    colorlinks=true,
    linkcolor=linkblue,
    citecolor=linkblue,
    urlcolor=linkblue
]{hyperref}

\usepackage{orcidlink}

\newcommand{\msun}{M_{\odot}}

\begin{document}

\title{An EOS-Driven Extension of \texttt{NSCool} for Compact Star Cooling with Hadronic and Quark Degrees of Freedom}

\author{%
Federico Nola\textsuperscript{1,2,3,*}\,
\orcidlink{0000-0001-9249-9547}
\qquad
Fernando Arias-Aragón\textsuperscript{4}\,
\orcidlink{0000-0002-1161-840X}
\\[1.2em]
\small
\textsuperscript{1}Universit\`a degli Studi della Campania ``Luigi Vanvitelli'',\\
Viale Abramo Lincoln 5, 81100 Caserta, Italy
\\[0.35em]
\textsuperscript{2}Istituto Nazionale di Fisica Nucleare, Sezione di Napoli,\\
Strada Comunale Cinthia, 80126 Napoli, Italy
\\[0.35em]
\textsuperscript{3}Istituto Nazionale di Fisica Nucleare, Laboratori Nazionali di Frascati,\\
Via Enrico Fermi 54, 00044 Frascati (RM), Italy
\\[0.35em]
\textsuperscript{4}Departamento de Geología, Física y Química Inorgánica, Universidad Rey Juan Carlos,\\
Calle Tulipán s/n, 28933 Móstoles, Madrid, Spain
\\[0.8em]
\small
\textsuperscript{*}Corresponding author:
\href{mailto:federico.nola@unicampania.it}
{\texttt{federico.nola@unicampania.it}}
}

\date{}
\maketitle

\begin{abstract}
We present an EOS-driven extension of the \texttt{NSCool} thermal evolution code that enables complete tabulated equations of state to be treated within a single composition-based input structure. The generalized \texttt{NEW} interface includes additional baryonic fractions, a bosonic composition variable, the hadronic volume fraction, and quark fractions, allowing nucleonic, hyperonic, resonant-baryonic, and hybrid hadron--quark configurations to be handled within the same workflow.

The neutrino sector is extended accordingly, including updated nucleonic modified Urca and bremsstrahlung rates, additional baryonic direct Urca channels, hyperonic processes, baryonic pair breaking and formation, and quark direct Urca, modified Urca, bremsstrahlung, and PBF contributions. In mixed phases, emissivities and the corresponding core heat capacity contributions are evaluated from phase-local quantities and combined using the hadronic volume fraction.

The implementation is validated against the original \texttt{NSCool} calculation for a controlled nucleonic benchmark and demonstrated with representative hadronic and quark-containing EOSs. These calculations are intended as software validation tests of the generalized workflow rather than as observational fits or statistical EOS inference.
\end{abstract}

\noindent\textbf{Keywords:} NSCool; neutron stars; compact stars; hybrid stars; thermal evolution; equation of state; neutrino emission

\section{Introduction}

The thermal evolution of compact stars provides a direct probe of the microscopic degrees of freedom that populate matter at supranuclear density. After the early proto-neutron star stage, the cooling of an isolated neutron star is governed by its stellar structure, thermal profile and local microphysics, which together determine how energy is transported and lost.

During the neutrino cooling era, weak processes in the core are especially sensitive to the particle content of dense matter, to the opening of fast direct Urca channels and to the suppression or enhancement induced by baryonic or quark pairing. Cooling calculations therefore provide a complementary tool to mass--radius measurements and tidal deformability constraints in the study of the dense matter equation of state (EOS) (\cite{YakovlevPethick2004,PageReddy2006,Potekhin2015}).

The connection between cooling and composition is especially important because the internal structure of compact stars is still uncertain. The simplest description assumes cold, catalyzed matter composed of neutrons, protons, electrons and muons in $\beta$ equilibrium. At the densities reached in the inner core, however, additional degrees of freedom may become energetically favored, modifying both the EOS and the neutrino emissivity.

Hyperons may open additional baryonic direct Urca channels, providing efficient cooling mechanisms even when the nucleonic direct Urca process is kinematically forbidden. Deconfined quarks can also contribute to rapid cooling through quark direct Urca, modified Urca, bremsstrahlung and, when paired, pair breaking and formation processes (\cite{Iwamoto1980,Iwamoto1982}).

\texttt{NSCool} is one of the standard public tools for simulating the thermal evolution of neutron stars (\cite{Page2016NSCool}). It solves the general relativistic thermal evolution equations for a spherically symmetric star and provides modular interfaces for the stellar structure, EOS input, envelope model and microphysical processes. Among its EOS options, the \texttt{NEW} framework is particularly useful for tabulated, composition-dependent hadronic EOSs, since the local particle fractions and effective masses can be supplied directly by the input table. In the original implementation, however, quark-containing configurations are treated through a separate workflow, rather than as part of the same tabulated composition structure.

This work develops an extension of the \texttt{NEW} framework in which complete tabulated EOSs containing hadronic and quark degrees of freedom can be handled within a unified input format\footnote{A patch that implements the changes presented here to the public version of \texttt{NSCool} is available at~\cite{nola_2026_21938691}.}. The extension preserves the EOS-driven logic of the original code, but enlarges the set of local quantities read from the EOS table by including additional baryonic fractions, a bosonic composition variable, the hadronic volume fraction and quark fractions. In this way, nucleonic, hyperonic, resonant-baryonic and hybrid configurations can be treated as different limits of the same composition-driven workflow.

The neutrino emission sector is also updated and generalized consistently with this EOS-driven approach. For nucleonic matter, the modified Urca and nucleon--nucleon bremsstrahlung rates are implemented following modern reanalyses of the standard \cite{FrimanMaxwell1979} framework (\cite{BottaroCaputoFiorillo2024}). The baryonic sector includes the implemented nucleonic channels, octet-hyperon channels and explicitly added $\Xi$ transitions, while the quark sector includes quark direct Urca, modified Urca, bremsstrahlung and pair breaking and formation contributions. The aim is not to derive new microscopic emissivities, but to embed established rates into a unified \texttt{NSCool} implementation in which the local composition of a complete EOS determines the active cooling channels.

This paper is organized as follows. Secs.~\ref{sec:nscool_original}--\ref{sec:implementation} describe the original \texttt{NSCool} treatment, the generalized EOS format and its implementation in the extended \texttt{NEW} workflow. Sec.~\ref{sec:neutrino_rates} summarizes the neutrino emission channels included in the updated framework, and Sec.~\ref{sec:discussion} discusses the scope and limitations of the implementation. Numerical checks are separated from the methodological development: Appendix~\ref{app:numerical_validation} documents the direct comparison with the original code, while Appendix~\ref{app:eos_examples} collects representative composition-driven EOS runs.

\section{Original \texttt{NSCool} Treatment of Tabulated EOSs and Quark Matter}
\label{sec:nscool_original}

A central feature of \texttt{NSCool} is its modular structure, which separates the stellar model, microphysical inputs, boundary conditions and heating prescriptions into independent routines. This organization makes it possible to modify selected components of the microphysics without altering the full thermal evolution solver. Among the available EOS input options, the \texttt{NEW} format provides the most flexible interface for tabulated, composition-dependent hadronic EOSs.

In the original \texttt{NEW} option, the EOS table supplies the thermodynamic quantities, lepton fractions, baryonic particle fractions and effective masses required by the cooling routines. These quantities are interpolated onto the stellar grid and used to activate the relevant hadronic neutrino processes according to the local composition, pairing inputs and kinematic thresholds. The particle content is therefore provided directly by the EOS table, rather than reconstructed from an analytic model.

Quark matter, however, is not handled within the same \texttt{NEW} infrastructure. In the original code, quark-containing configurations are treated through a separate quark-specific branch, associated with the quark matter EOS option and with its own input quantities. While useful for dedicated quark star or quark core calculations, this separation prevents nucleonic, hyperonic and hybrid configurations from being treated as different limits of a single complete EOS-driven workflow.

This becomes restrictive for complete EOS tables generated by modern microscopic or phenomenological models, where different particle species and phases may appear at different densities within the same stellar model. In such cases, the cooling calculation should follow the local composition continuously from a single EOS table, activating the appropriate hadronic or quark processes according to the local particle fractions and phase content. This is the main technical motivation for the generalized EOS-driven extension described in the following sections.

\section{Generalized EOS Format}
\label{sec:eos_format}

The extension described in this work is based on a generalized tabulated EOS format designed to provide the cooling routines with the local thermodynamic and compositional information required for nucleonic, hyperonic, resonant-baryonic and hybrid hadron--quark matter. Starting from the original \texttt{NEW} structure, which includes thermodynamic variables, lepton fractions, selected baryonic fractions and baryonic effective masses, we add the quantities needed to describe complete EOS tables with deconfined quarks and extended mixed phases.

The original input structure may be schematically written as
\begin{equation*}
\left\{\rho,
P,
n_B,
Y_e,
Y_\mu,
Y_n,
Y_p,
Y_\Lambda,
Y_{\Sigma^-},
Y_{\Sigma^0},
Y_{\Sigma^+},
m_p^\ast,
m_n^\ast,
m_\Lambda^\ast,
m_{\Sigma^-}^\ast,
m_{\Sigma^0}^\ast,
m_{\Sigma^+}^\ast\right\},
\label{eq:old_eos_format}
\end{equation*}
where $\rho$ is the mass density, $P$ is the pressure, $n_B$ is the baryon number density, $Y_i$ denotes the number fraction of particle species $i$, and $m_i^\ast$ is the corresponding effective mass. The extended input table keeps these quantities and adds new columns describing further baryonic species, possible bosonic contributions, the local hadronic volume fraction and the quark composition.

The generalized format used in the present work can be written schematically as
\begin{align*}
\mathrm{EOS}_{\rm ext}=\Big\{
&\rho,
P,
n_B,
Y_e,
Y_\mu,
Y_B,
m_B^\ast,
Y_{\rm extra},
m_{\rm extra}^\ast,
f_{\rm had},
Y_{\rm boson},
Y_u,
Y_d,
Y_s
\Big\},
\label{eq:extended_eos_format}
\end{align*}
where $Y_B$ and $m_B^\ast$ denote the baryonic fractions and effective masses already present in the original \texttt{NEW} format, while $Y_{\rm extra}$ and $m_{\rm extra}^\ast$ denote the additional baryonic degrees of freedom introduced in the extended table. In the implementation considered here, these additional components include $\Xi^-$ and $\Xi^0$ hyperons, the $\Delta^-$, $\Delta^0$, $\Delta^+$, and $\Delta^{++}$ resonances, the excited strange baryons $\Sigma^{\ast -}$, $\Sigma^{\ast 0}$, $\Sigma^{\ast +}$, $\Xi^{\ast -}$, $\Xi^{\ast 0}$ and the $\Omega^-$ baryon. The quantity $Y_{\rm boson}$ is the global bosonic contribution to the baryon number budget supplied by the EOS table, normalized to the total baryon density $n_B$. The corresponding density used internally in the consistency control is $n_{\rm boson}=Y_{\rm boson}n_B$. In the present implementation this variable contributes to the baryon number consistency check, but it is not assigned an electric charge term and does not by itself activate any neutrino emission channels.

The last three entries, $Y_u$, $Y_d$, and $Y_s$, describe the up, down and strange quark fractions. The extended table therefore contains more degrees of freedom than those for which explicit neutrino emission processes are currently implemented. In particular, the decuplet baryons are propagated as part of the EOS composition, but they are not treated as $\beta$ decaying species in the present cooling implementation. Consequently, no processes involving $\Delta$, $\Sigma^\ast$, $\Xi^\ast$, or $\Omega^-$ states are added. The active baryonic weak channels are restricted to the octet baryon transitions listed in Sec.~\ref{sec:neutrino_rates}, supplemented by the explicitly included $\Xi$ octet channels.

The role of $f_{\rm had}$ is central in the treatment of hybrid configurations. It specifies the local fraction of the volume occupied by hadronic matter. Therefore,
\begin{equation*}
0 \leq f_{\rm had} \leq 1,
\end{equation*}
with $f_{\rm had}=1$ corresponding to purely hadronic matter and $f_{\rm had}=0$ corresponding to purely quark matter. Intermediate values describe a mixed hadron--quark region, as obtained for example in EOSs with an extended mixed phase. The quark volume fraction is consequently
\begin{equation*}
f_{\rm q}=1-f_{\rm had}.
\label{eq:fquark}
\end{equation*}
This prescription makes the format flexible enough to describe both sharp and continuous transitions.

The particle fractions supplied by the EOS table are interpreted as global fractions normalized to the total baryon density \(n_B\). In mixed hadron--quark regions, however, the microscopic rates must be evaluated using densities reconstructed inside the phase in which the corresponding particles propagate. The explicit phase-local reconstruction used in the implementation is given in Sec.~\ref{sec:implementation}.

The local densities obtained in this way determine the Fermi momenta entering the neutrino rates. For baryons and leptons, the convention used in the code is
\[
k_{F,i}=\left(3\pi^2 n_i\right)^{1/3},
\]
whereas for a quark flavor $f=u,d,s$ the color degeneracy is included through
\[
k_{F,f}=\left(\pi^2 n_f^{\rm q}\right)^{1/3}.
\]
Here the densities are the appropriate local densities in the corresponding phase. In practice, the implementation enforces the limiting cases \(f_{\rm had}=1\) for purely hadronic regions and \(f_{\rm had}=0\) for purely quark regions, avoiding spurious divisions by zero.

The quark fractions also determine the baryon number and electric charge carried by the quark component. Since each quark carries baryon number $1/3$, the baryonic contribution of the quark sector is
\begin{equation*}
Y_B^{\rm q}=\frac{1}{3}\left(Y_u+Y_d+Y_s\right).
\label{eq:quark_baryon_fraction}
\end{equation*}
Similarly, using the quark electric charges $q_u=2/3$ and $q_d=q_s=-1/3$, the quark charge fraction is
\begin{equation*}
Y_Q^{\rm q}=\frac{1}{3}\left(2Y_u-Y_d-Y_s\right).
\label{eq:quark_charge_fraction}
\end{equation*}
Here $Y_{\rm boson}$ enters the baryon-number consistency condition as
\begin{equation}
\sum_B Y_B+\frac{Y_u+Y_d+Y_s}{3}+Y_{\rm boson}=1.
\end{equation}

These relations provide useful consistency checks for EOS tables containing quark matter. In particular, when the EOS is in beta equilibrium and charge neutral, the lepton fractions and the charged hadronic and quark fractions must combine to give vanishing total charge within the numerical accuracy of the table.

A key feature of the generalized format is that nucleonic, hyperonic, resonant-baryonic and hybrid configurations are represented within the same table structure. Purely nucleonic matter is recovered by setting all additional baryonic, bosonic and quark entries to zero and imposing \(f_{\rm had}=1\). Hyperonic matter is described by nonzero hyperon fractions at \(f_{\rm had}=1\), while hybrid configurations are obtained by allowing nonzero quark fractions and assigning \(f_{\rm had}\) according to the phase structure of the EOS.

The local neutrino emissivity is then determined by the tabulated composition and phase content. Purely hadronic regions activate only hadronic channels, purely quark regions activate only quark channels and mixed regions are treated through the phase-weighting prescription described in Sec.~\ref{sec:implementation}. In this way, different classes of compact star matter are selected directly by the EOS table, without switching between independent cooling workflows. This construction also preserves the standard hadronic limit when the additional columns vanish.

\section{Implementation in the Extended \texttt{NEW} Framework}
\label{sec:implementation}

The implementation follows the modular structure of \texttt{NSCool} and modifies the EOS composition interface together with selected core microphysics routines. The thermal evolution solver, stellar structure input, time integration and envelope treatment are left unchanged. The extended \texttt{NEW} workflow reads and interpolates the additional composition variables, reconstructs phase-local densities where needed and supplies the resulting local quantities to the neutrino emissivity and specific-heat routines.

In the extended implementation, the \texttt{NEW} reader is modified to read the additional composition and phase variables defined in Sec.~\ref{sec:eos_format} from the same tabulated EOS as the standard inputs. The full local composition can then be propagated consistently to the cooling routines without switching to a separate quark matter workflow.

The interpolation strategy is kept close to the original \texttt{NSCool} prescription. At each radial point of the stellar model, the quantities read from the EOS table are interpolated between neighboring density points. If $X$ denotes any thermodynamic or composition variable, one may write schematically
\begin{equation*}
X(r_i)=w_1 X(\rho_1)
+
w_2 X(\rho_2),
\label{eq:implementation_interpolation}
\end{equation*}
where the weights are computed from the logarithmic density interval containing the local density $\rho(r_i)$. The same procedure is applied to all additional composition and phase variables introduced in the extended EOS format.

After interpolation, the tabulated global fractions are converted into the densities entering the microscopic rates. In purely hadronic regions, $f_{\rm had}=1$ and the hadronic densities reduce to $n_i=Y_i n_B$, while the quark contribution is absent. In purely quark regions, $f_{\rm had}=0$ and only quark processes are retained. In mixed regions, the EOS fractions are interpreted as global fractions normalized to the total baryon density, and the rates are evaluated using phase-local densities. Thus, for a hadronic species $i$ one uses

\begin{equation}
n_i^{\rm had}(r)=\frac{Y_i(r)n_B(r)}{f_{\rm had}(r)},
\qquad
f_{\rm had}(r)>0,
\label{eq:local_hadronic_density}
\end{equation}
while for a quark flavor $f=u,d,s$ one uses
\begin{equation}
n_f^{\rm q}(r)=\frac{Y_f(r)n_B(r)}{1-f_{\rm had}(r)},
\qquad
f_{\rm had}(r)<1.
\label{eq:local_quark_density}
\end{equation}
The corresponding Fermi momenta, kinematic thresholds, and pairing suppressions are then computed from these phase-local densities.

The neutrino emission routines receive the interpolated local composition, effective masses, phase fractions  and pairing inputs and activate the corresponding hadronic or quark channels according to the thresholds and suppression factors described in Sec.~\ref{sec:neutrino_rates}. Thus, nucleonic, hyperonic and quark contributions are handled within the same extended \texttt{NEW} workflow.

Decuplet baryons included in the extended EOS table are treated as composition variables only. In the present implementation, the $\Delta$, $\Sigma^\ast$, $\Xi^\ast$, and $\Omega^-$ fractions do not activate additional direct Urca, modified Urca, bremsstrahlung or PBF channels. They can still affect the cooling indirectly by modifying the local composition, Fermi momenta, effective masses  and threshold conditions of the active nucleonic and hyperonic processes.

At each radial point, the code evaluates only the emissivities associated with the nonzero local particle fractions. A process is inactive if one of the required species has vanishing fraction, if the corresponding Fermi momentum is zero, or if the relevant kinematic threshold is not satisfied. The local hadronic and quark contributions are then combined according to
\begin{equation}
Q_\nu(r,T)=f_{\rm had}(r)
Q_\nu^{\rm had}(r,T)
+
\left[1-f_{\rm had}(r)\right]
Q_\nu^{\rm q}(r,T).
\label{eq:implementation_phase_weighting}
\end{equation}
This expression applies to the local volume emissivity. The thermal solver then integrates the resulting emissivity over the stellar profile in the usual \texttt{NSCool} way. In the limit $f_{\rm had}=1$, Eq.~\eqref{eq:implementation_phase_weighting} reduces to the ordinary hadronic emissivity, whereas for $f_{\rm had}=0$ it reduces to the pure quark contribution; intermediate values describe mixed hadron--quark matter. The same volume fraction prescription is applied to the core heat capacity terms included in the implementation, with the nucleonic, $\Lambda$ and $\Sigma$ contributions weighted by $f_{\rm had}$ and the quark contribution by $1-f_{\rm had}$. No additional heat capacity terms for $\Xi$ or decuplet baryons are included in the present release.

Pairing effects are included through the same general mechanism used by \texttt{NSCool}: each emissivity kernel is multiplied by the appropriate reduction factor when one or more participating species are paired. For baryons, the existing singlet and triplet pairing controls are retained. The source contains singlet PBF kernels for $\Sigma^-$, $\Sigma^+$, $\Xi^0$, and $\Xi^-$, but this release does not provide critical temperature models for these four species; their corresponding $T_c$ arrays therefore remain at the default floor unless source level gap profiles are supplied. For quarks, the rate and heat capacity routines retain explicit color resolved $T_c$ arrays. The quark pairing input used by this release constructs one flavor dependent critical temperature for each of $u$, $d$, and $s$ and assigns that value to all three color entries of the corresponding flavor. The direct Urca, modified Urca, bremsstrahlung and PBF quark rates are then suppressed or activated according to the local ratio $T/T_c$ for the relevant flavor and color. This allows the same EOS to be evolved under different assumptions about quark superconductivity without modifying the EOS table itself.

With this prescription, different stellar compositions are selected directly by the local EOS entries. The extended \texttt{NEW} workflow reduces to the standard hadronic limit for $f_{\rm had}=1$, describes hyperonic matter when hyperon fractions are present, and activates quark emissivities locally wherever quark fractions and a nonzero quark volume fraction are supplied by the EOS table.

\section{Neutrino Emission Rates}
\label{sec:neutrino_rates}

The generalized EOS format described above must be supplemented by a local emissivity library. In this work we do not derive new microscopic rates. Instead, we reorganize established neutrino emissivities into a single composition-driven structure that can be evaluated locally from the tabulated particle fractions, effective masses, Fermi momenta, pairing inputs and phase content. The local emissivity is decomposed into hadronic and quark contributions,
\begin{equation*}
Q_\nu^{\rm had}
=
Q_\nu^{N}+Q_\nu^{Y},
\qquad
Q_\nu^{q}=Q_\nu^{\rm quark},
\label{eq:qnu_total_components}
\end{equation*}
where $Q_\nu^{N}$ denotes nucleonic processes, $Q_\nu^{Y}$ denotes the implemented hyperonic and additional $\Xi$ channels and $Q_\nu^{q}$ denotes quark processes. The purpose of this section is to summarize the channels retained in the updated implementation and the minimal emissivity structure used by the code. More detailed derivations of the individual kernels can be found in the references cited below.

\subsection{Nucleonic Modified Urca and Bremsstrahlung}
\label{subsec:nucleonic_slow}

For nucleonic matter, the slow neutrino emission sector contains modified Urca and nucleon--nucleon bremsstrahlung reactions. The modified Urca branches are
\begin{align*}
n+n &\rightarrow n+p+\ell+\bar\nu_\ell,
&
n+p+\ell &\rightarrow n+n+\nu_\ell,\\
n+p &\rightarrow p+p+\ell+\bar\nu_\ell,
&
p+p+\ell &\rightarrow n+p+\nu_\ell,
\end{align*}
with $\ell=e,\mu$ when the corresponding lepton branch is present and kinematically allowed. The bremsstrahlung channels are
\begin{equation*}
n+n\rightarrow n+n+\nu+\bar\nu,
\qquad
n+p\rightarrow n+p+\nu+\bar\nu,
\qquad
p+p\rightarrow p+p+\nu+\bar\nu.
\end{equation*}
The unpaired modified Urca and bremsstrahlung kernels are updated following the Bottaro--Caputo--Fiorillo reanalysis of the Friman--Maxwell rates \citep{FrimanMaxwell1979,BottaroCaputoFiorillo2024}. In our implementation they are multiplied by the high density suppression factor used by \citet{Buschmann2022} and by the usual superfluid reduction factors:
\begin{equation*}
    \begin{aligned}
Q_{{\rm MU},n}^{\rm impl}
&=
R_{{\rm MU},n}^{\rm SF}\,\gamma^6\,Q_{{\rm MU},n}^{\rm BCF},\\
Q_{{\rm MU},p}^{\rm impl}
&=
R_{{\rm MU},p}^{\rm SF}\,\gamma^6\,Q_{{\rm MU},p}^{\rm BCF},\\
Q_{{\rm Br},ij}^{\rm impl}
&=
R_{{\rm Br},ij}^{\rm SF}\,\gamma^6\,Q_{{\rm Br},ij}^{\rm BCF},
\end{aligned} \qquad
ij=nn,np,pp .
\label{eq:nucleonic_slow_impl}
\end{equation*}

The suppression factor is
\begin{equation*}
\gamma=
\left[
1+\frac{1}{3}m_{np}^{\ast}
\left(
\frac{k_{F,n}}{1.68\,{\rm fm}^{-1}}
\right)
\right]^{-1},
\qquad
m_{np}^{\ast}=\frac{m_n^{\ast}+m_p^{\ast}}{2}.
\label{eq:gamma_suppression}
\end{equation*}
This prescription is applied only to the slow nucleonic modified Urca and bremsstrahlung channels. The nucleonic direct Urca process, when allowed by momentum conservation, remains part of the standard direct Urca sector.

\subsection{Baryonic Direct Urca Processes}

Fast cooling is possible when a direct Urca process satisfies the Fermi momentum triangle condition. For a generic baryonic transition,
\begin{equation*}
B_1\rightarrow B_2+\ell+\bar\nu_\ell,
\qquad
B_2+\ell\rightarrow B_1+\nu_\ell,
\end{equation*}
the process is active only if
\begin{equation*}
k_{F,B_1}\leq k_{F,B_2}+k_{F,\ell},
\label{eq:durca_triangle}
\end{equation*}
with the analogous inverse condition. The implemented baryonic direct Urca emissivity is written schematically as
\begin{equation*}
Q_{\rm DU}^{B_1B_2}
=
r_{B_1B_2}
\left(4.24\times10^{27}\right)
 m_{B_1}^{\ast}m_{B_2}^{\ast}
T_9^6 L_\ell R_{\rm DU}^{B_1B_2},
\label{eq:baryonic_durca_compact}
\end{equation*}
where $r_{B_1B_2}$ contains the weak coupling coefficient, $L_\ell$ is the lepton phase-space factor, and $R_{\rm DU}^{B_1B_2}$ is the pairing reduction factor. The standard implemented channels are
\begin{equation*}
n\rightarrow p,
\qquad
\Lambda\rightarrow p,
\qquad
\Sigma^-\rightarrow n,
\qquad
\Sigma^-\rightarrow\Lambda,
\qquad
\Sigma^-\rightarrow\Sigma^0.
\end{equation*}
The updated implementation adds the octet $\Xi$ transitions
\begin{equation*}
\Xi^-\rightarrow\Lambda,
\qquad
\Xi^-\rightarrow\Sigma^0,
\qquad
\Xi^0\rightarrow\Sigma^+,
\qquad
\Xi^-\rightarrow\Xi^0.
\end{equation*}
These channels are activated locally by the EOS composition and by the corresponding threshold condition.

\subsection{Hyperonic Modified Urca and Bremsstrahlung}

Hyperons also contribute to slow cooling through modified Urca-like and baryon--baryon bremsstrahlung reactions. The implemented hyperonic slow rates follow the structure of the \cite{Maxwell1987} treatment of neutrino processes in hyperonic matter. A generic hyperonic modified Urca process has the form
\begin{equation*}
B_1+B_2\rightarrow B_3+B_4+\ell+\bar\nu_\ell,
\end{equation*}
whereas a generic hyperonic bremsstrahlung process is
\begin{equation*}
B_1+B_2\rightarrow B_3+B_4+\nu+\bar\nu.
\end{equation*}
Both classes scale as $T^8$. In compact notation the implemented rates are
\begin{align*}
Q_{\rm MU}^{Y}
&=
\sum_a R_a^{\rm MU}\,C_a^{\rm MU}\,
I_a^{\rm MU}(k_{F,1},k_{F,2},k_{F,3},k_{F,4})\,T^8,
\\
Q_{\rm Br}^{Y}
&=
\sum_a R_a^{\rm Br}\,C_a^{\rm Br}\,
I_a^{\rm Br}(k_{F,1},k_{F,2},k_{F,3},k_{F,4})\,T^8.
\end{align*}
The implemented hyperonic modified Urca channels are
\begin{equation*}
\Sigma^-\Sigma^-\rightarrow\Sigma^-\Lambda,
\quad
\Sigma^-\Lambda\rightarrow\Lambda\Lambda,
\quad
\Sigma^- n\rightarrow\Sigma^- p,
\quad
\Sigma^- n\rightarrow\Lambda n,
\quad
\Sigma^- p\rightarrow\Lambda p,
\end{equation*}
with the charged lepton and antineutrino emitted in the weak vertex. The implemented hyperonic bremsstrahlung channels are
\begin{equation*}
\Sigma^- n\rightarrow\Sigma^- n,
\quad
\Sigma^- p\rightarrow\Sigma^- p,
\quad
\Sigma^-\Lambda\rightarrow\Sigma^-\Lambda,
\quad
\Sigma^-\Sigma^-\rightarrow\Sigma^-\Sigma^-,
\quad
\Sigma^-p\rightarrow\Lambda n,
\end{equation*}
with a neutrino--antineutrino pair in the final state. The channel coefficients and momentum-space integrals are those of the cited microscopic treatment.

\subsection{Fast Exotic and Kaon}

The fast exotic sector already present in the original code is retained. In particular, the kaon condensate emissivity is kept in the form
\begin{equation*}
Q_K=
\frac{5}{4}\sin^2\theta_K\sin^2\theta_C
\left(2.21\times10^{26}\right)
 m_n^{\ast}m_p^{\ast}
\left(\frac{\mu_e}{100\,{\rm MeV}}\right)
(1+3g_A^2)T_9^6,
\label{eq:kaon_emissivity}
\end{equation*}
and the phenomenological fast contribution is written as
\begin{equation*}
Q_{\rm exo}=C_{\rm exo}
\left(\frac{\rho}{2.8\times10^{14}\,{\rm g\,cm^{-3}}}\right)^{2/3}
T_9^{p_{\rm exo}},
\qquad
Q_{\rm fast}=Q_K+Q_{\rm exo}.
\end{equation*}
The legacy kaon kernel remains in the source, but the generalized \texttt{NEW} reader initializes its kaon angle $\theta_K$ to zero and does not map $Y_{\rm boson}$ onto that variable. Consequently, the new bosonic column does not activate kaon emission. No new kaonic, pionic, or boson-induced neutrino channel is introduced by $Y_{\rm boson}$.

\subsection{PBF Contributions}

For baryonic pair breaking and formation, the updated implementation keeps the standard singlet and triplet nucleonic channels and contains singlet-state hyperonic PBF kernels for $\Sigma^-$, $\Sigma^+$, $\Xi^0$, and $\Xi^-$. As noted in Sec.~\ref{sec:implementation}, this release does not expose dedicated $T_c$ models for these four species through the distributed pairing input, so these terms are inactive with the default floor values unless corresponding source-level gap profiles are supplied. The baryonic PBF contribution implemented at kernel level is summarized as
\begin{equation*}
Q_{\rm PBF}^{B}
=
Q_{\rm PBF}^{n\,{}^1S_0}
+Q_{\rm PBF}^{p\,{}^1S_0}
+Q_{\rm PBF}^{n\,{}^3P_2}
+
\sum_{Y=\Sigma^-,\Sigma^+,\Xi^0,\Xi^-}
Q_{\rm PBF}^{Y\,{}^1S_0}.
\label{eq:baryon_pbf_total}
\end{equation*}
The PBF rate is multiplied by the corresponding high density factor as
\begin{equation*}
Q_{\rm PBF}^{B,{\rm impl}}=\gamma_B^2 Q_{\rm PBF}^{B,0},
\label{eq:pbf_gamma2}
\end{equation*}
while vector-current corrections and the axial-dominated singlet-channel treatment follow the standard PBF literature \citep{YakovlevKaminkerLevenfish1999,KaminkerHaenselYakovlev1999,LeinsonPerez2006}.

\subsection{Quark Emission Channels}

For deconfined quark matter, the direct Urca reactions are
\begin{equation*}
d\rightarrow u+\ell+\bar\nu_\ell,
\qquad
s\rightarrow u+\ell+\bar\nu_\ell,
\label{eq:quark_durca_reactions}
\end{equation*}
together with the corresponding inverse reactions. The charged lepton is $\ell=e,\mu$ when the relevant branch is present; there is no independent sum over three neutrino flavors in these charged-current processes. The direct Urca contribution is
\begin{equation*}
Q_{\rm DU}^{q}=Q_{\rm DU}^{ud}+Q_{\rm DU}^{us},
\label{eq:qdurca_total}
\end{equation*}
with unpaired kernels of the Iwamoto type \citep{Iwamoto1980,Iwamoto1982}. Color-resolved pairing suppression is retained through three factors for each branch,
\begin{align*}
Q_{\rm DU}^{ud}
&=
\left(R_{ud}^{(1)}+R_{ud}^{(2)}+R_{ud}^{(3)}\right)
Q_{{\rm DU},0}^{ud},\\
Q_{\rm DU}^{us}
&=
\left(R_{us}^{(1)}+R_{us}^{(2)}+R_{us}^{(3)}\right)
Q_{{\rm DU},0}^{us}.
\end{align*}
Thus, in the unpaired limit the explicit color sum gives the factor three associated with quark color and no additional factor of three is applied to quark direct Urca.

The quark modified Urca and bremsstrahlung kernels are implemented as compact Iwamoto-type parametrizations \citep{Iwamoto1982}. In this release they are evaluated from the phase-local quark baryon density
\begin{equation*}
n_B^{q}=\frac{n_u^{q}+n_d^{q}+n_s^{q}}{3},
\end{equation*}
and take the form
\begin{align*}
Q_{\rm MU}^{q}
&=
R_{\rm MU}^{q}
\left[2.83\times10^{19}\alpha_c^2
\left(\frac{n_B^{q}}{0.16\,{\rm fm}^{-3}}\right)T_9^8\right]
\;{\rm erg\,cm^{-3}\,s^{-1}},\\
Q_{\rm Br}^{q}
&=
R_{\rm Br}^{q}
\left[2.98\times10^{19}
\left(\frac{n_B^{q}}{0.16\,{\rm fm}^{-3}}\right)T_9^8\right]
\;{\rm erg\,cm^{-3}\,s^{-1}}.
\end{align*}
The factors $R_{\rm MU}^{q}$ and $R_{\rm Br}^{q}$ are averages over the active $ud$ and $us$ branches and over the three color slots within each active branch; they approach unity in the unpaired limit, so no additional overall factor of three multiplies these two compact slow emission kernels.

The quark PBF contribution is evaluated as an explicit sum over flavors and colors,
\begin{equation*}
Q_{\rm PBF}^{q}
=
\sum_{f=u,d,s}
\sum_{c=1}^{3}
\Theta\!\left(T_{c,f}^{(c)}-T\right)
A_q
\left(\frac{k_{F,f}}{1.68\,{\rm fm}^{-1}}\right)
T_9^7
F_q\!\left(v_f^{(c)}\right),
\label{eq:quark_pbf_general}
\end{equation*}
with $A_q=3\times10^{21}$ and $v_f^{(c)}=\Delta_f^{(c)}(T)/T$. This term should be understood as a phenomenological singlet-pairing PBF contribution motivated by studies of neutrino emission in paired quark matter \citep{JaikumarPrakash2001,JaikumarRobertsSedrakian2006}.

In the original code, several strange quark and color-superconducting contributions are grouped in a monolithic routine. In the updated implementation, the quark sector is reorganized as
\begin{equation*}
Q_\nu^{q}
=
Q_{\rm DU}^{q}
+Q_{\rm MU}^{q}
+Q_{\rm Br}^{q}
+Q_{\rm PBF}^{q},
\label{eq:qnu_quark_total}
\end{equation*}
so that each term can be activated locally according to the quark fractions, phase content, Fermi momenta and critical temperatures supplied by the pairing input. In this release the three color entries of a given flavor receive the same flavor dependent $T_c$, while the rate formulas retain the explicit color sums.

\subsection{Phase Weighting and Activation of Processes}

At each radial point, the active neutrino channels are selected by the local EOS composition, the corresponding kinematic thresholds, and the pairing suppressions discussed above. In mixed hadron--quark regions, the hadronic and quark contributions are combined through the local phase fractions. The total local emissivity is written as
\begin{align}
Q_\nu(r,T)
&=
f_{\rm had}(r)
\left[
Q_{\rm DU}^{B}
+Q_{\rm MU}^{B}
+Q_{\rm Br}^{B}
+Q_{\rm PBF}^{B}
+Q_{\rm fast}
\right]
\nonumber\\
&\quad+
\left[1-f_{\rm had}(r)\right]
\left[
Q_{\rm DU}^{q}
+Q_{\rm MU}^{q}
+Q_{\rm Br}^{q}
+Q_{\rm PBF}^{q}
\right].
\label{eq:total_emissivity_final}
\end{align}
The baryonic contribution contains the implemented nucleonic channels, octet-hyperon channels and explicitly added $\Xi$ transitions, while the quark contribution is activated only where the EOS supplies nonzero quark fractions and a nonzero quark volume fraction.

The microscopic rates entering Eq.~\eqref{eq:total_emissivity_final} are evaluated using the phase-local densities reconstructed from the global EOS fractions, as defined in Eqs.~\eqref{eq:local_hadronic_density}--\eqref{eq:local_quark_density}. This prescription preserves the original hadronic limit and activates quark emission only in regions where a quark component is explicitly present in the EOS table.

\section{Discussion and conclusions}
\label{sec:discussion}
\label{sec:conclusions}

This work introduces a composition-driven extension of the \texttt{NSCool} \texttt{NEW} workflow for complete tabulated EOSs. Nucleonic, octet-hyperonic and quark emissivities are selected from the local tabulated composition, while phase-local densities and a hadronic volume fraction allow hadronic, quark and mixed regions to be handled by one interface. The contribution is methodological: established emissivities are reorganized rather than rederived.

The generalized table is intentionally broader than the present emissivity library. Decuplet baryons and possible bosonic components can be propagated as EOS variables, but they do not activate dedicated channels in this version. Their effect is indirect, through the stellar structure, particle fractions, Fermi momenta, effective masses and thresholds of the implemented nucleonic, octet-hyperonic, $\Xi$ and quark processes.

The phase-weighted prescription provides a consistent computational interpolation between the purely hadronic and purely quark limits when the EOS supplies an unambiguous local volume fraction. Its microscopic interpretation remains tied to the thermodynamic construction of the input EOS and to whether its tabulated fractions are global or phase local.

The numerical validation of the updated implementation is presented in Appendices~\ref{app:numerical_validation} and~\ref{app:eos_examples}. The former compares the original and updated codes under controlled nucleonic conditions, while the latter demonstrates the generalized workflow for representative hadronic and quark-containing EOS tables. These calculations are intended solely as software validation tests and not as observational fits or EOS inference.

When the additional exotic and quark fractions vanish and $f_{\rm had}=1$, the generalized implementation reduces to the standard hadronic case. The extended framework thus preserves backward compatibility while providing a flexible basis for future studies of composition dependent compact star cooling, including different pairing scenarios, quark critical temperatures and mixed phase prescriptions.

\section*{Acknowledgments}

The authors acknowledge the MUSES Collaboration and the developers of the MUSES Calculation Engine for providing the modular computational framework used to construct and process the representative dense matter EOSs employed in this work.

\section*{Data and software availability}

The original \texttt{NSCool} package is available from its official distribution site \citep{Page2016NSCool}. The extension developed in this work is publicly available at \texttt{\url{https://doi.org/10.5281/zenodo.21938691}} as the portable \texttt{NSCool Update 1.0.0} patch. The patch is designed to be applied to the compatible original distribution and provides the updated source files together with the tools and instructions required for installation and execution. The EOS tables distributed with the official \texttt{NSCool} package are automatically converted to a format compatible with the extended \texttt{NEW} interface when the patch is applied. The original \texttt{NSCool} package is not redistributed.

The MUSES CE EOS tables used for the representative calculations in this work are available from the authors upon reasonable request and can also be independently reproduced using the publicly available MUSES Calculation Engine \texttt{\url{https://musesframework.io}}, following the same EOS construction workflow and model choices adopted here. The DS(CMF) EOS tables used in the analysis are publicly available through the CompOSE database.

\bibliographystyle{plainnat}
\bibliography{references}

@misc{nola_2026_21938691,
  author       = {Nola, Federico and
                  Arias-Aragón, Fernando},
  title        = {EOS-Driven Extension for Hadronic and Quark
                   Compact Star Cooling: NSCool Extension Tool
                  },
  month        = aug,
  year         = 2026,
  publisher    = {Zenodo},
  doi          = {10.5281/zenodo.21938691},
  url          = {https://doi.org/10.5281/zenodo.21938691},
  swhid        = {swh:1:dir:03e7e380460b2201bd72733fcca80f98be5a0725
                   ;origin=https://doi.org/10.5281/zenodo.21938690;vi
                   sit=swh:1:snp:7c0f92f17e5f11350c65c2701e1b098d22d4
                   7e47;anchor=swh:1:rel:8f63963444f912bc91bc94fe0f7e
                   1eb8b6473997;path=NSCool\_Update\_1.0.0\_Patch
                  },
}

@article{YakovlevPethick2004,
  author        = {Yakovlev, D. G. and Pethick, C. J.},
  title         = {Neutron Star Cooling},
  journal       = {Annual Review of Astronomy and Astrophysics},
  volume        = {42},
  pages         = {169--210},
  year          = {2004},
  doi           = {10.1146/annurev.astro.42.053102.134013},
  eprint        = {astro-ph/0402143},
  archivePrefix = {arXiv}
}

@article{PageReddy2006,
  author        = {Page, Dany and Reddy, Sanjay},
  title         = {Dense Matter in Compact Stars: Theoretical Developments and Observational Constraints},
  journal       = {Annual Review of Nuclear and Particle Science},
  volume        = {56},
  pages         = {327--374},
  year          = {2006},
  doi           = {10.1146/annurev.nucl.56.080805.140600},
  eprint        = {astro-ph/0608360},
  archivePrefix = {arXiv}
}

@article{Potekhin2015,
  author        = {Potekhin, A. Y. and Pons, J. A. and Page, Dany},
  title         = {Neutron Stars---Cooling and Transport},
  journal       = {Space Science Reviews},
  volume        = {191},
  pages         = {239--291},
  year          = {2015},
  doi           = {10.1007/s11214-015-0180-9},
  eprint        = {1507.06186},
  archivePrefix = {arXiv},
  primaryClass  = {astro-ph.HE}
}

@misc{Page2016NSCool,
  author       = {Page, Dany},
  title        = {{NSCool}: Neutron Star Cooling Code},
  year         = {2016},
  publisher    = {Astrophysics Source Code Library},
  version      = {ascl:1609.009},
  url          = {https://www.astroscu.unam.mx/neutrones/NSCool/}
}

@article{FrimanMaxwell1979,
  author  = {Friman, B. L. and Maxwell, O. V.},
  title   = {Neutrino Emissivities of Neutron Stars},
  journal = {The Astrophysical Journal},
  volume  = {232},
  pages   = {541--557},
  year    = {1979},
  doi     = {10.1086/157313}
}

@article{BottaroCaputoFiorillo2024,
  author        = {Bottaro, Salvatore and Caputo, Andrea and Fiorillo, Damiano F. G.},
  title         = {Neutrino Emission in Cold Neutron Stars: Bremsstrahlung and Modified Urca Rates Reexamined},
  journal       = {Journal of Cosmology and Astroparticle Physics},
  volume        = {2024},
  number        = {11},
  pages         = {015},
  year          = {2024},
  doi           = {10.1088/1475-7516/2024/11/015},
  eprint        = {2406.18640},
  archivePrefix = {arXiv},
  primaryClass  = {hep-ph}
}

@article{Maxwell1987,
  author  = {Maxwell, O. V.},
  title   = {Neutrino Emissivities of Hyperon Stars},
  journal = {The Astrophysical Journal},
  volume  = {316},
  pages   = {691--707},
  year    = {1987},
  doi     = {10.1086/165234}
}

@article{Iwamoto1980,
  author  = {Iwamoto, Naoki},
  title   = {Quark Beta Decay and the Cooling of Neutron Stars},
  journal = {Physical Review Letters},
  volume  = {44},
  pages   = {1637--1640},
  year    = {1980},
  doi     = {10.1103/PhysRevLett.44.1637}
}

@article{Iwamoto1982,
  author  = {Iwamoto, Naoki},
  title   = {Neutrino Emissivities and Mean Free Paths of Degenerate Quark Matter},
  journal = {Annals of Physics},
  volume  = {141},
  pages   = {1--49},
  year    = {1982},
  doi     = {10.1016/0003-4916(82)90271-8}
}

@article{Buschmann2022,
  author        = {Buschmann, Malte and Dessert, Christopher and Foster, Joshua W. and Long, Andrew J. and Safdi, Benjamin R.},
  title         = {Upper Limit on the QCD Axion Mass from Isolated Neutron Star Cooling},
  journal       = {Physical Review Letters},
  volume        = {128},
  pages         = {091102},
  year          = {2022},
  doi           = {10.1103/PhysRevLett.128.091102},
  eprint        = {2111.09892},
  archivePrefix = {arXiv},
  primaryClass  = {hep-ph}
}

@article{YakovlevKaminkerLevenfish1999,
  author        = {Yakovlev, D. G. and Kaminker, A. D. and Levenfish, K. P.},
  title         = {Neutrino Emission Due to Cooper Pairing of Nucleons in Cooling Neutron Stars},
  journal       = {Astronomy and Astrophysics},
  volume        = {343},
  pages         = {650--660},
  year          = {1999},
  eprint        = {astro-ph/9812366},
  archivePrefix = {arXiv}
}

@article{KaminkerHaenselYakovlev1999,
  author        = {Kaminker, A. D. and Haensel, P. and Yakovlev, D. G.},
  title         = {Neutrino Emission Due to Proton Pairing in Neutron Stars},
  journal       = {Astronomy and Astrophysics},
  volume        = {345},
  pages         = {L14--L16},
  year          = {1999},
  eprint        = {astro-ph/9904166},
  archivePrefix = {arXiv}
}

@article{LeinsonPerez2006,
  author        = {Leinson, L. B. and Perez, A.},
  title         = {Vector Current Conservation and Neutrino Emission from Singlet-Paired Baryons in Neutron Stars},
  journal       = {Physics Letters B},
  volume        = {638},
  pages         = {114--118},
  year          = {2006},
  doi           = {10.1016/j.physletb.2006.05.036},
  eprint        = {astro-ph/0606651},
  archivePrefix = {arXiv}
}

@article{JaikumarPrakash2001,
  author        = {Jaikumar, Prashanth and Prakash, Madappa},
  title         = {Neutrino Pair Emission from Cooper Pair Breaking and Recombination in Superfluid Quark Matter},
  journal       = {Physics Letters B},
  volume        = {516},
  pages         = {345--352},
  year          = {2001},
  doi           = {10.1016/S0370-2693(01)00945-4},
  eprint        = {astro-ph/0105225},
  archivePrefix = {arXiv}
}

@article{JaikumarRobertsSedrakian2006,
  author        = {Jaikumar, Prashanth and Roberts, Craig D. and Sedrakian, Armen},
  title         = {Direct Urca Neutrino Rate in Colour Superconducting Quark Matter},
  journal       = {Physical Review C},
  volume        = {73},
  pages         = {042801},
  year          = {2006},
  doi           = {10.1103/PhysRevC.73.042801},
  eprint        = {nucl-th/0509093},
  archivePrefix = {arXiv}
}

@article{AriasAragonNola2026,
  title = {Core composition effects on the QCD axion mass limit from neutron star cooling},
  author = {Arias-Aragón, Fernando and Nola, Federico},
  journal = {Phys. Rev. D},
  pages = {},
  year = {2026},
  month = {Aug},
  publisher = {American Physical Society},
  doi = {10.1103/3frv-1wm7},
}

@article{Pelicer2025MUSES,
  author        = {Reinke Pelicer, Mateus and Cruz-Camacho, Nikolas and Conde, Carlos and Friedenberg, David and Roy, Satyajit and Zhang, Ziyuan and Manning, T. Andrew and Alford, Mark G. and Clevinger, Alexander and Grefa, Joaquin and others},
  title         = {Building Neutron Stars with the MUSES Calculation Engine},
  journal       = {Physical Review D},
  volume        = {111},
  pages         = {103037},
  year          = {2025},
  doi           = {10.1103/PhysRevD.111.103037},
  eprint        = {2502.07902},
  archivePrefix = {arXiv},
  primaryClass  = {nucl-th}
}

@article{Pelicer2025MUSESOverview,
  author        = {Pelicer, Mateus and Dexheimer, Veronica and Grefa, Joaquin},
  title         = {An Overview of the MUSES Calculation Engine and How It Can Be Used to Describe Neutron Stars},
  eprint        = {2505.14921},
  archivePrefix = {arXiv},
  primaryClass  = {nucl-th},
  year          = {2025}
}

@misc{MusesCalculationEngine,
author = {Manning, T. Andrew},
doi = {10.5281/zenodo.14721911},
month = {2},
title = {MUSES Calculation Engine},
url = {https://musesframework.io},
year = {2025}
}

@article{DexheimerSchramm2008,
author        = {Dexheimer, V. and Schramm, S.},
title         = {Proto-Neutron and Neutron Stars in a Chiral {SU(3)} Model},
journal       = {Astrophysical Journal},
volume        = {683},
pages         = {943--948},
year          = {2008},
eprint        = {0802.1999},
archivePrefix = {arXiv},
primaryClass  = {astro-ph},
doi           = {10.1086/589735}
}

@article{DexheimerSchramm2010,
author        = {Dexheimer, V. A. and Schramm, S.},
title         = {A Novel Approach to Model Hybrid Stars},
journal       = {Physical Review C},
volume        = {81},
pages         = {045201},
year          = {2010},
eprint        = {0901.1748},
archivePrefix = {arXiv},
primaryClass  = {astro-ph.SR},
doi           = {10.1103/PhysRevC.81.045201}
}

@article{Dexheimer2017,
author        = {Dexheimer, V.},
title         = {Tabulated Neutron Star Equations of State Modeled within the Chiral Mean Field Model},
journal       = {Publications of the Astronomical Society of Australia},
volume        = {34},
pages         = {e006},
year          = {2017},
eprint        = {1708.08342},
archivePrefix = {arXiv},
primaryClass  = {astro-ph.HE}
}

@article{DexheimerGomesKlahnHanSalinas2021,
author        = {Dexheimer, V. and Gomes, R. O. and Klähn, T. and Han, S. and Salinas, M.},
title         = {{GW190814} as a Massive Rapidly Rotating Neutron Star with Exotic Degrees of Freedom},
journal       = {Physical Review C},
volume        = {103},
number        = {2},
pages         = {025808},
year          = {2021},
eprint        = {2007.08493},
archivePrefix = {arXiv},
primaryClass  = {astro-ph.HE},
doi           = {10.1103/PhysRevC.103.025808}
}

\newpage
\FloatBarrier
\appendix

\renewcommand{\thefigure}{\thesection.\arabic{figure}}
\renewcommand{\thetable}{\thesection.\arabic{table}}
\setcounter{figure}{0}
\setcounter{table}{0}

\section{Numerical validation of \texttt{NSCool Update 1.0.0}}
\label{app:numerical_validation}

\begin{figure}[ht]
\centering
\includegraphics[width=0.5\textwidth]{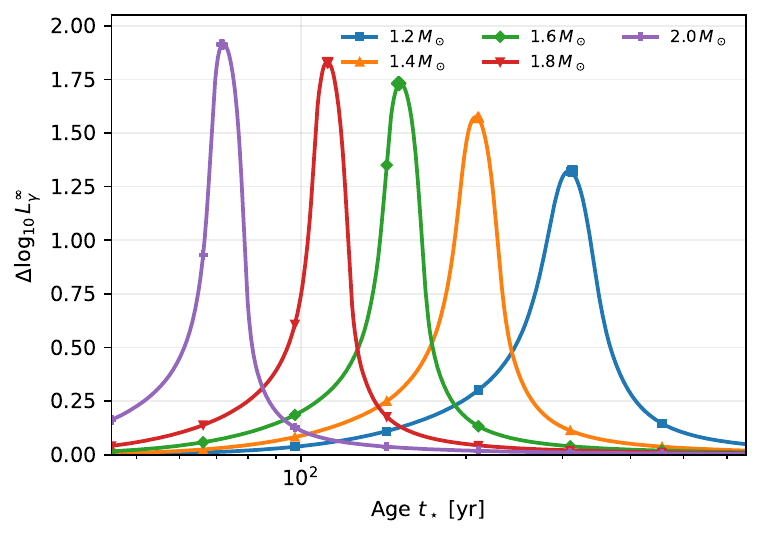}
\caption{Difference in photon luminosity between the updated and original \texttt{NSCool} implementations for the MUSES CE $npe\mu$ sequence. Large filled symbols indicate the maximum deviations within the rapid cooling transition. The $1.0\,\msun$ sequence is omitted because its maximum occurs near the late-time cutoff.}
\label{fig:delta_npemu}
\end{figure}

This appendix documents the numerical comparison of \texttt{NSCool Update 1.0.0} with the original \texttt{NSCool} implementation. The calculations are verification tests of the released computational update and are not used for observational likelihood analysis, population inference or EOS selection.

All calculations use TOV profiles generated from the selected EOS, an iron envelope, vanishing pairing gaps, standard conductivity and no heating, magnetic field or accretion. PBF is therefore inactive in these runs and the cooling curves are deliberately unsuppressed software validation baselines rather than calibrated population models.

\subsection{Direct comparison with the original \texttt{NSCool} implementation}
\label{subsec:comparison}

The controlled comparison uses the MUSES CE $npe\mu$ EOS and gravitational masses $M=1.0,1.2,1.4,1.6,1.8$ and $2.0\,\msun$. The EOS, stellar profiles and all auxiliary inputs are identical in the original and updated calculations. Exotic fractions vanish and $f_{\rm had}=1$, so the hyperonic, resonant-baryonic and quark additions are inactive.

The active physical difference is the simultaneous replacement of both nucleonic modified Urca branches and the $nn$, $np$ and $pp$ bremsstrahlung kernels by the prescriptions described in Sec.~\ref{subsec:nucleonic_slow}, including the $\gamma^6$ high density factor. Nucleonic direct Urca is unchanged, and PBF is absent because all gaps vanish. The comparison therefore tests the complete revised slow nucleonic sector, not modified Urca alone.

To isolate the narrow transition displacement we use the code comparison diagnostic
\begin{equation}
 \Delta\log_{10}L_\gamma^\infty(t;M)
 =\log_{10}\!\left[
 \frac{L_{\gamma,{\rm updated}}^\infty(t;M)}
      {L_{\gamma,{\rm original}}^\infty(t;M)}
 \right].
\label{eq:delta_lgamma}
\end{equation}
Positive values correspond to a larger photon luminosity in the updated run at the same age, while negative values correspond to a smaller luminosity.

The largest deviations are concentrated in the narrow interval where the cooling curves cross the steep luminosity drop at slightly different ages. Because the curves are nearly vertical in this region on a log--log cooling diagram, a modest horizontal displacement can generate a large instantaneous luminosity ratio without implying a global renormalization of the cooling curve. The $1.0\,\msun$ sequence is a special late-time case: its largest ratio occurs close to the final luminosity cutoff, where the small denominator amplifies a minor timing difference.

Figure~\ref{fig:delta_npemu} shows only the transition window for $1.2$--$2.0\,\msun$. The $1.0\,\msun$ maximum is excluded from the figure because it occurs at the final late-time cutoff, where the small denominator amplifies a minor timing difference. For $1.2$--$2.0\,\msun$, the maximum $\Delta\log_{10}L_\gamma^\infty$ ranges from $1.32$ to $1.91$ at ages decreasing from about $313$ to $72$ yr. Over the full common computed domains the most negative deviation is only $-5.2\times10^{-3}$. The update therefore shifts the rapid transition rather than globally rescaling the luminosity.

\setcounter{figure}{0}
\setcounter{table}{0}
\section{Representative EOS workflow checks}
\label{app:eos_examples}

\begin{figure}[ht]
\centering
\includegraphics[width=0.85\linewidth]{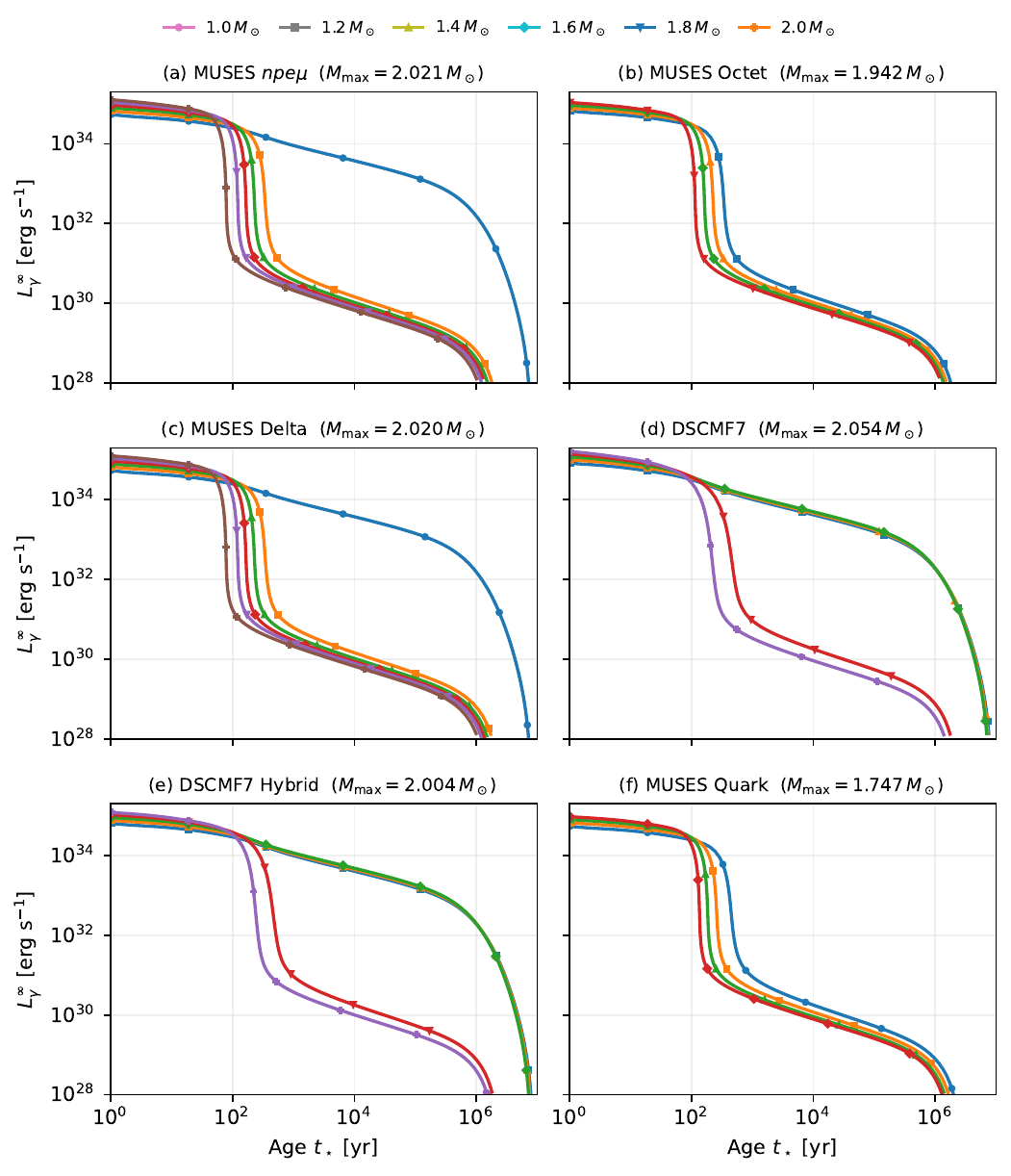}
\caption{Representative cooling sequences for six EOS tables using the same auxiliary inputs. Only stable configurations are shown. The panels illustrate the generalized workflow across hadronic and quark-containing compositions and are not intended as EOS fits or observational comparisons.}
\label{fig:eos_grid}
\end{figure}

This appendix illustrates the generalized composition-driven interface using representative EOS compositions under the same auxiliary setup adopted for the numerical validation. The calculations are intended as functional tests of the workflow rather than as EOS comparisons or fits to the observed cooling population.

The representative EOS set is taken from Ref.~\cite{AriasAragonNola2026}. The nucleonic, octet, delta and quark-containing MUSES CE models were generated with the MUSES Calculation Engine \citep{MusesCalculationEngine,Pelicer2025MUSES,Pelicer2025MUSESOverview}; DSCMF7 and DSCMF7 Hybrid provide external DS(CMF) benchmarks \citep{DexheimerSchramm2008,DexheimerSchramm2010,Dexheimer2017,DexheimerGomesKlahnHanSalinas2021}.

The same interface is then exercised on the MUSES CE $npe\mu$, Octet, Delta and Quark EOSs and the DSCMF7 and DSCMF7 Hybrid tables. Figure~\ref{fig:eos_grid} collects the representative stable lines retained from the validated calculations. The production maximum masses are shown in the panel titles; configurations above those maxima are not plotted.

More massive MUSES $npe\mu$, Octet and Delta configurations generally reach the rapid luminosity drop earlier because their cores sample higher densities and larger volumes of efficient emission. Delta states affect these calculations only indirectly because the current library assigns them no dedicated weak channel. The DSCMF7 and DSCMF7 Hybrid panels remain close over much of the adopted baseline setup, while the most massive cases enter rapid cooling earlier. The MUSES Quark panel demonstrates local activation of quark emissivities from the same table structure. These curves establish functionality across different EOS compositions; a unique physical attribution would require radial activation volumes and process-by-process ablation runs.

\end{document}